\documentclass[journal]{IEEEtran}
\usepackage{cite}

\ifCLASSINFOpdf
  \usepackage[pdftex]{graphicx}
  \DeclareGraphicsExtensions{.pdf,.jpeg,.png}
\else
\fi
\usepackage[cmex10]{amsmath}
\usepackage{array}

\usepackage{mdwmath}
\usepackage{mdwtab}

\usepackage[]{caption}
\usepackage[font=footnotesize]{subfig}
\usepackage{url}

\begin{document}
%
\title{A Spin-Light Polarimeter for Multi-GeV Longitudinally Polarized Electron Beams}
%
%
%

\author{Prajwal~Mohanmurthy,
        and~Dipangkar~Dutta
\thanks{P. Mohanmurthy and D. Dutta are with the Department of Physics and Astronomy at Mississippi State University, Mississippi State, MS 39762.}
\thanks{Manuscript received July 11, 2013}}

%
%

\markboth{Journal of \LaTeX\ Class Files,~Vol.~6, No.~1, January~2007}%
{Shell \MakeLowercase{\textit{et al.}}: Bare Demo of IEEEtran.cls for Journals}
%



\maketitle

\begin{abstract}
The physics program at the upgraded Jefferson Lab (JLab) and the physics program envisioned for the proposed electron-ion collider (EIC) include large efforts to search for interactions beyond the Standard Model (SM) using parity violation in electroweak interactions. These experiments require precision electron polarimetry with an uncertainty of $<$ 0.5 \%. The spin dependent Synchrotron radiation (SR), called ''spin-light,'' can be used to monitor the electron beam polarization. In this article we develop a conceptual design for a ``spin-light'' polarimeter that can be used at a high intensity, multi-GeV electron accelerator. We have also built a Geant4 based simulation for a prototype device and report some of the results from these simulations.

\end{abstract}


%
\IEEEpeerreviewmaketitle

\section{Introduction}
%
%
%
%
The determination of the longitudinal polarization of the electron beam is one of the dominant systematic uncertainties in any parity violating electron scattering (PVES) experiment. In order to achieve the desired high precision, the polarization of the electron beam must be monitored continuously with an uncertainty of $<$0.5\%. These ambitious goals can be achieved if multiple independent and high precision polarimeters are used simultaneously. In addition to being precise, the polarimeters must be non-invasive and must achieve the desired statistical precision in the shortest time possible. Compton and M{\o}ller polarimeters are typically the polarimeters of choice for these experiments and are essential to achieve the desired precision. However, a complimentary polarimetry technique based on the spin dependence of synchrotron radiation, referred to as ``spin-light,'' can be used as a relative polarimeter. A spin-light polarimeter could provide additional means for improving the systematic uncertainties and when calibrated against a Compton/M{\o}ller polarimeter it could provide a stable continuous monitoring of the beam polarization. We develop the conceptual design for a continuous polarimeter based on ``spin-light''. The proposed spin-light polarimeter can achieve statistical precision of $<$ 1\% in measurement cycles of less than 10 minutes for 4 - 20 GeV electron beams with a beam currents of $\sim$ 100 $\mu$A. 

\begin{table}
\caption{A comparison of the Compton, M{\o}ller and Spin-light polarimeters.}
\label{tab:polcomp}
\begin{center}
\begin{tabular}{|l|l|l|}\hline
Compton & Spin-Light & M{\o}ller \\ \hline
non-invasive,& non-invasive,    & invasive \\
continuous & continuous  &     \\\hline
analyzing power & analyzing power & analyzing power \\
energy dependent& energy dependent& energy independent \\ \hline
high currents & moderately & low currents \\ 
              &   high currents &               \\\hline
target is 100\% & no target & target is $<$ 10\% \\
 polarized  &  needed   &   polarized \\
(requires stable laser)   &         &  \\\hline
electron \& photon &  beam left \& right & no independent \\
detection are &   detectors provide & measurements \\
two  independent &two  independent &  possible \\     
measurements&  measurements & \\\hline 
high precision & high precision & high precision \\
absolute polarimeter &  relative polarimeter & absolute polarimeter \\ \hline
Best reported~\cite{sld} & expected & Best reported~\cite{hallcmoller} \\
instrumental & instrumental & instrumental \\
uncertainty: 0.4\%  & uncertainty: 0.6\%& uncertainty: 0.47\% \\\hline
Best reported~\cite{sld} & estimated & Best achieved~\cite{qweak}\\ 
absolute & absolute & absolute \\
uncertainty: 0.5\%  & uncertainty: $\sim$5\% & uncertainty: 0.85\% \\\hline
 
\end{tabular}
\end{center}
\end{table}

M{\o}ller and Compton polarimeters have a proven track record of very high precision, the JLab Hall-C M{\o}ller polarimeter has an instrumental uncertainty of 0.47\%~\cite{hallcmoller} and absolute uncertainty of 0.85\%~\cite{qweak}, while the Compton polarimeter used in the SLD experiment achieved an instrumental uncertainty of 0.4\%~\cite{sld} and an absolute uncertainty of 0.5\%~\cite{sld}, hence they are essential for any PVES program. However, a spin-light polarimeter would have a few operational and instrumental advantages over conventional polarimeters, such that when used in parallel with Compton/M{\o}ller polarimeters they might help reduce the systematic uncertainties and achieve the very high precision essential for the future PVES program. For example, M{\o}ller polarimeters use a polarized Fe target, and the polarization of the target is difficult to determine and may depend on the beam intensity.  Moreover, M{\o}ller polarimeters operate at low current, and are invasive to the primary experiment. Compton polarimeters require a stable laser (the photon target) and are very sensitive to backgrounds.  The proposed spin-light polarimeter is a target free device, hence it should be easier to 
operate over long periods, with its stability governed just by the stability of the electron beam. Moreover, this novel polarimeter would facilitate cross-checks and systematic studies when used with other conventional polarimeters. On the other hand, one of the disadvantages is that the proposed device can achieve comparable instrumental uncertainties only as a relative polarimeter, whereas the absolute polarization is 
what is required in the PVES experiments. Nevertheless a precise and stable relative polarimeter can be a very useful device. The spin-light polarimeter could be used in conjunction with a Compton polarimeter, such that the difficult to operate Compton  polarimeter is used for calibration and the easier to operate and stable spin-light polarimeter is used to continuous monitor the beam polarization. Moreover, only the M{\o}ller and the spin-light polarimeters allow measurement of the transverse component of a longitudinally polarized electron beam. The key features of conventional polarimeters and a spin-light polarimeter are summarized in Table~\ref{tab:polcomp}.

In this article we discuss the theory behind the spin-light polarimeter and present a complete conceptual design for such a polarimeter. In order to study some of the systematic uncertainties of the novel spin-light polarimeter we have built a Geant4~\cite{geant4} based simulation of the polarimeter. The results from these simulations are also presented here. This work was inspired by a 1993 proposal by Karabekov and Rossmanith~\cite{karabekov93}.

\section{Synchrotron Radiation}
At the typical energies and magnetic fields of present day terrestrial accelerators, the intensity and angular distribution of SR from an electron moving along a curved path, under the influence of a magnetic field, is described with high precision by classical electrodynamics~\cite{classical}. The total radiation rate for highly relativistic electrons, $P^{clas}$ is $\propto E^4$ as given by the Larmor formula: $P^{clas} = \frac{2}{3}\frac{e^2\gamma^4 c}{R^2}$, where $e$ is the electron charge, $v_e$ is the velocity of the electron, $c$ is the velocity of light in vacuum, $\beta = \frac{v_e}{c}$, $\gamma = \frac{1}{\sqrt{1-\beta^2}} = \frac{E}{m_e c^2}$ is the Lorentz boost, $m_e$ is the rest mass of the electron, and, $R$ is the radius of curvature of the electron orbit. The angular distribution of the radiated power is given by  
\begin{equation}
\frac{dP}{d\Omega} = \frac{e^2\gamma^4 c}{4\pi R^2}\frac{(1-\beta \cos \theta)^2 - (1 - \beta^2)\sin^2 \theta \cos^2 \phi}{(1 - \beta \cos \theta)^5},
\end{equation}

where the angles $(\theta , \phi)$ are measured with respect to the direction of the electron's motion. For highly relativistic electrons ($\gamma >> 1$), the radiation is Lorentz boosted in the forward direction with an opening angle $\theta \approx 1/\gamma$. The classical theory also tells us that SR is strongly linearly polarized~\cite{polar}, with $P_{\sigma} = \frac{7}{8}P$ and $P_{\pi}=\frac{1}{8}P$, where $\sigma$-component is the one where the electric field lies in the plane of the electron orbit while $\pi$-component is the one where the electric field lies in the plane perpendicular to the orbital plane. Experimentally, these properties of SR have been demonstrated with high precision for a wide range of frequencies ~\cite{sync1} - \cite{sync4}.

Classical theory also tells us that the cone of SR passes over a fixed angle $\theta$ in retarded~\footnote{accounts for the finite delay between the photon emission and detection} time $\Delta t^{'} \approx \frac{\Delta \theta}{\omega_0} = \frac{1}{\gamma \omega_0}$, where $\omega_0 = v_e/R \approx c/R$ is the Larmor frequency for the electron. For a distant observer the corresponding time interval is $\Delta t = \Delta t^{'} ( 1 - \beta) \approx 1/(2\gamma^3 \omega_0)$. Hence the spectral width of the radiation is $\Delta \omega \approx 1/\Delta t = 2 \gamma^3 \omega_{0}$. This implies that the radiation is strong at high harmonics of the Larmor frequency $(\omega_0)$ and can be considered as continuous. However from quantum mechanics we know that the radiation at frequency $\omega$ consists of photons of energy $\hbar \omega$. Thus there must exist a sufficiently strong acceleration such that a single photon will carry away all of the electron's energy ($\gamma m_e c^2 = \hbar \omega_c$). The critical magnetic field strong enough to provide this acceleration is found to be $B_{crit} = \frac{ m^2 c^3}{e\hbar}$ = 4.41$\times$10$^{9}$ Tesla. The frequency of the photon radiated under the influence of $B_{crit}$ is called the critical frequency ($\omega_c$) and is given by$\omega_c = \frac{3}{2} \gamma^3 \omega_0 = \frac{3}{2} \gamma^3 c/R $. From this we get the critical energy $E_{crit} = m_e c^2\sqrt{\frac{m_e c R}{\hbar}} \approx $ 10$^{6}$ GeV. The extremely large values of the critical field and critical energy help explain why the classical theory is successful at the energies and fields accessible at present day accelerators. However, it turns out that several quantum effects appear at considerably lower electron energies.

\subsection{Quantum Theory of Synchrotron Radiation}
 The exact expression for SR intensity including quantum corrections was calculated by Sokolov, Ternov and Klepikov, based on the solution to the Dirac equation in the framework of quantum electrodynamics~\cite{stk52}. They showed that at energies above a few 100 MeV, there would be fluctuations in the radius of the electron orbit leading to radial oscillations~\cite{sokolov53} of the electron trajectory and quantum widening of the trajectory similar to Brownian motion. These oscillations and widening of the trajectory are essential in determining the dynamics of electrons in an accelerator, specially storage rings. Sokolov and Ternov also developed the mathematics required to describe the spin of relativistic electrons moving in an external electromagnetic field~\cite{sttext,ternov95}, which allowed them to calculate the electron spin related properties of SR. 

The power radiated by electrons undergoing transitions $n \rightarrow n^{'}$ (related to the radius of the electron orbit), $s \rightarrow s^{'}$ (quadratic fluctuation  of the radius) and $ j \rightarrow j^{'}$ (spin orientations with respect to the magnetic field), integrated over angles and summed over polarization states, is given by~\cite{sttext,ternov95}:
\begin{equation} P = P^{clas} \times \frac{9 \sqrt{3}}{16 \pi} \sum_{s^{'}} \int_{0}^{\infty} \frac{y dy}{(1 + \xi y)^4}I^2_{ss^{'}}(x)F(y),
\end{equation} 
where, 
\begin{IEEEeqnarray}{rCl}
 F(y) &=& \frac{1+ jj^{'}}{2}[ 2(1+\xi y)\int_{y}^{\infty}K_{5/3}(x)dx \nonumber \\
      &+& \frac{1}{2}\xi^2 y^2 K_{2/3}(y) - j(2+\xi y)\xi y K_{1/3}(y)] \nonumber \\
      &+& \frac{1 - jj^{'}}{2}\xi^2 y^2 \left[K_{2/3}(y) + l K_{1/3}(y)\right]
\end{IEEEeqnarray}
 
where $P^{clas}$ is the classical expression for SR power radiated, $y=\frac{\omega}{\omega_c}$, $x = \frac{3}{4}\frac{\xi \gamma^3 y^2}{(1 + \xi y)^2}$, $I_{ss^{'}}$ are the Laguerre functions, $K_{n}(x)$ are modified Bessel functions, and $\xi = \frac{3}{2}\frac{B}{B_{crit}}\gamma$ is the critical parameter. Because, $B_{crit}$ = 4.41$\times$ 10$^{9}$ Tesla we have $\xi << 1$ for magnetic fields used at all man-made accelerators and the above expression can be expanded in terms of the critical parameter $\xi$  to get~\cite{bordo};
\begin{IEEEeqnarray}{rCl} 
P &=& P^{clas} [(1 -\frac{55 \sqrt{3}}{24}\xi + \frac{64}{3}\xi^2) \nonumber \\ 
  &-& (\frac{1+jj^{'}}{2})(j\xi + \frac{5}{9}\xi^2 \frac{245 \sqrt{3}}{48}j \xi^2)\nonumber \\ 
  &+& (\frac{1-jj^{'}}{2})(\frac{4}{3}\xi^2 + \frac{315\sqrt{3}}{432}j\xi^2) + ...]
\end{IEEEeqnarray}
These quantum corrections to the classical expression for charge radiation ($P^{clas}$) involve contributions from the electron recoil effects of radiation, interference of the charge radiation and radiation due to the intrinsic magnetic moment of the electron, magnetic moment radiation due to Larmor precession, magnetic moment radiation due to Thomas precession, interference between the Larmor and Thomas radiation and radiation due to the anomalous magnetic moment of an electron~\cite{bordo}. The lowest order spin-dependent correction is of order $\xi$ and the lowest order spin-flip correction term is of order $\xi^2$. The difference (of order $\xi$) between the expression for power radiated by polarized and unpolarized (spin averaged) electron beams has the form,
\begin{IEEEeqnarray}{rCl}
P^{pol} - P^{unpol} & = & -j\xi P^{clas} \int_{0}^{\infty} \frac{9 \sqrt{3}}{8 \pi}y^2 K_{1/3}(y) dy. 
\end{IEEEeqnarray}
The above expression is directly related to the spin polarization of an electron beam $j$, hence the difference $P^{pol} - P^{unpol} = P^{spin}$ can be called ``spin-light''~\cite{sttext,ternov95}. This offers a new possibility for visual or direct observation of the polarization characteristics of an electron beam by determining the SR power at a fixed range of spectral frequency.

The spin dependence of the SR was verified at the VEPP-4 storage ring in Novosibirsk~\cite{belom82}, using a 3 pole wiggler magnet (called a magnetic snake). The intensity of the SR produced by the wiggler for transversely polarized electrons was monitored while the beam was periodically depolarized using a RF field. The measured variation in SR intensity with polarization matched exactly with the expectation from the Sokolov-Ternov theory. This spin dependent part of the SR has been successfully used at the VEPP-4 to monitor the transverse polarization of the electron beam.

The spin-flip term has a unique effect on circular accelerators. The radiation probability with spin flip is given by $w^{\uparrow \downarrow} = \frac{1}{\tau}\left( 1 + j \frac{8\sqrt{3}}{15}\right)$, where $j=$ 1 is for spin along the magnetic field and $j=$ -1 is for spin opposite to the magnetic field~\cite{sttext}. Using this relation, Sokolov and Ternov had predicted that, over time the beam in a circular accelerator would eventually become polarized opposite to the direction of the magnetic field. In other words the beam of a circular accelerator becomes transversely polarized over time via self polarization. This phenomena known as the Sokolov-Ternov self polarization was first observed at the French storage ring~\cite{marin} at Orsay and is now routinely used to polarize beams at circular electron accelerators such as DESY.

\section{Spin Light}
\label{sec:spinlight}
In the discussion above the spin orientation $j$ is relative to the magnetic field that produces the SR, with $j=$ 1 along the magnetic field. Polarized electron beams have longitudinal ($p_z$) and transverse ($p_{\perp}$) components relative to the beam direction and the transverse polarization have vertical and horizontal components. For a vertically oriented magnetic field, the total SR power from transversely polarized electrons, ignoring spin flip terms and other terms of order $\xi^2$, is given by~\cite{sttext}:
\begin{IEEEeqnarray}{rCl}
 P_{\gamma}(tran) & = & \frac{9 n_e}{16 \pi^3}\frac{ce^2}{R^2}\gamma^5\int_{0}^{\infty}\frac{y^2dy}{(1+\xi y)^4}\oint d\Omega (1 + \alpha ^2)^2 \nonumber\\
&\times & [ K^2_{2/3}(z) + \frac{\alpha^2}{1 + \alpha^2}K^2_{1/3}(z) \nonumber \\ 
&-& p_{\perp}\xi y \frac{1}{\sqrt{1+\alpha^2}}K_{1/3}(z)K_{2/3}(z)],
\end{IEEEeqnarray}
where $n_e$ is the number of electrons, $z= \frac{\omega}{2\omega_c}(1 + \alpha^2)^{3/2}$, and $\alpha = \gamma \psi$, where $\psi$ is the vertical angle in the frame of the moving electron. The rest of the symbols are as defined in previous instances. The polarization dependent term in the above expression is an even function of the vertical angle therefore when integrated over all angles it makes the total SR power spin dependent. Thus by measuring this spin dependence in the total SR power radiated one can measure the transverse polarization of the electron beam. 

On the other hand, the total SR power from longitudinally polarized electrons, ignoring spin flip terms and other terms of order $\xi^2$, is given by~\cite{sttext}:  
\begin{IEEEeqnarray}{rCl}
 P_{\gamma}(long) & = & \frac{9 n_e}{16 \pi^3}\frac{ce^2}{R^2}\gamma^5\int_{0}^{\infty}\frac{y^2dy}{(1+\xi y)^4}\oint d\Omega (1 + \alpha ^2)^2 \nonumber\\
&\times & [K^2_{2/3}(z) + \frac{\alpha^2}{1 + \alpha^2}K^2_{1/3}(z) \nonumber \\ 
&+& p_z\xi y \frac{\alpha}{\sqrt{1+\alpha^2}}K_{1/3}(z)K_{2/3}(z)],
\end{IEEEeqnarray}
The spin dependent term in the above expression is an odd function of the vertical angle therefore when integrated over all angles it goes to zero and the total SR power for longitudinally polarized electrons is spin independent. However, the power radiated into the space above ($0 <\psi <\pi/2$) and below ($-\pi/2 <\psi <0$) the orbital plane of the electron are different and the difference between them is spin dependent. Therefore, by measuring this spatial asymmetry one can monitor the longitudinal polarization of the electron beam. 
For $\xi y << 1$, if we divide the above expression by the energy of the radiated photon, $E_{\gamma} = \frac{3}{2}\frac{\hbar c}{R} \gamma^3 y$ we get the total number of photons radiated into a finite horizontal angle $\Delta \theta$ as, 

\begin{figure}[hbt]
\centerline{\includegraphics*[width=9.0cm]{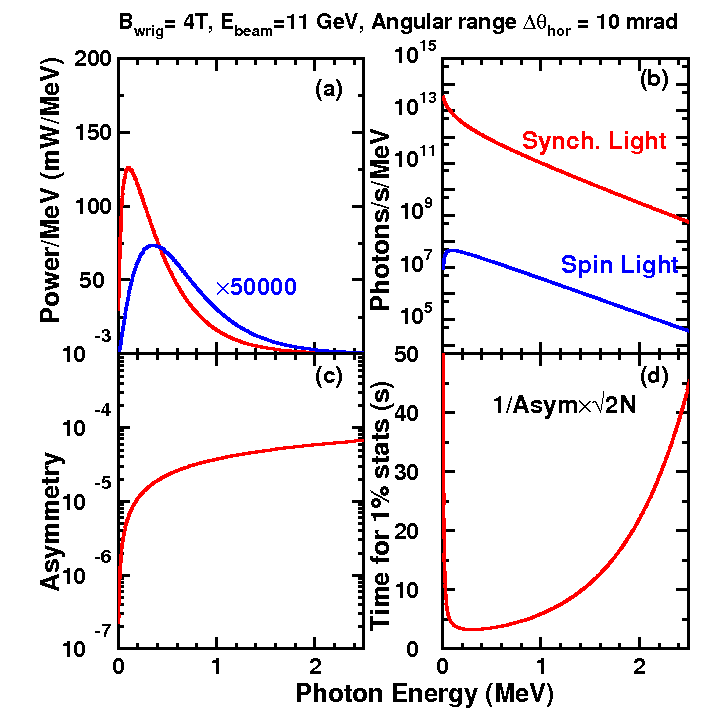}}
\caption[]{(a) The total SR power radiated, per MeV (red), and the spin dependent difference in power radiated, per MeV (blue), above and below the orbital plane, for 11 GeV, longitudinally polarized electrons in a 4~T magnetic field and 100 $\mu$A current, and 10\% detector efficiency. (b) The total number of SR photons per MeV and the number of ``spin-light'' photons per MeV, and (c) the asymmetry $\frac{\Delta N}{N}$ as a function of photon energy. (d) The time for 1\% statistical uncertainty, in seconds.}
\label{fig3}	
\end{figure}

\begin{IEEEeqnarray}{rCl}
N_{\gamma} &=& \frac{3}{4\pi^2}\frac{1}{137}\frac{I_e}{e}\gamma \Delta \theta \int_{y_1}^{y_2} ydy \int_{-\alpha}^{\alpha}(1+\alpha^2)^{3/2} \nonumber \\
&\times&\left[K^{2}_{2/3}(z) + \frac{\alpha^2}{1+\alpha^2}K^{2}_{1/3}(z)\right]d\alpha,
\end{IEEEeqnarray}
where $I_e$ is the beam current. The difference in the photon flux radiated in the the space above and below the electron orbit is given by,
\begin{IEEEeqnarray}{rCl}
\Delta N_{\gamma}(p_z) &=& \frac{3}{\pi^2}\frac{1}{137}\frac{I_e}{e} p_z \xi \gamma \Delta \theta \int_{y_1}^{y_2} y^2 dy \int_{0}^{\alpha}\alpha (1+\alpha^2)^{3/2}\nonumber \\
&\times& K_{1/3}(z) K_{2/3}(z) d\alpha
\end{IEEEeqnarray}

To examine the size and characteristics of the spin dependence we have numerically integrated the above two expressions for longitudinally polarized electron with 100\% polarization, in a 4 Tesla magnetic field, with $I_e$ = 100 $\mu$A, and $E_e$ = 11 GeV. We have integrated over a horizontal angular acceptance of $\Delta \theta$ = 10 mrad, and a vertical acceptance of $\alpha = \pm$1. The characteristic spectra of SR and spin-light obtained from these numerical intragrations are shown in Fig.~\ref{fig3}. The total power radiated $P_{\gamma}$(long) and  the spin dependent difference of power radiated above and below the orbital plane of the electron $\Delta P$(long) are shown as a function of photon energy in Fig~\ref{fig3}~(a). The number of SR photons $N_{\gamma}$(long) and the number  of spin-light photons $\Delta N_{\gamma}$(long), as function of photon energy are shown in Fig~\ref{fig3}~(b). The asymmetry defined as $A = \frac{\Delta N_{\gamma}(long)}{N_{\gamma}(long)}$ as a function of photon energy is shown in Fig.~\ref{fig3}~(c). 
\begin{figure}[hbt]
\centerline{{\includegraphics*[width=4.5cm]{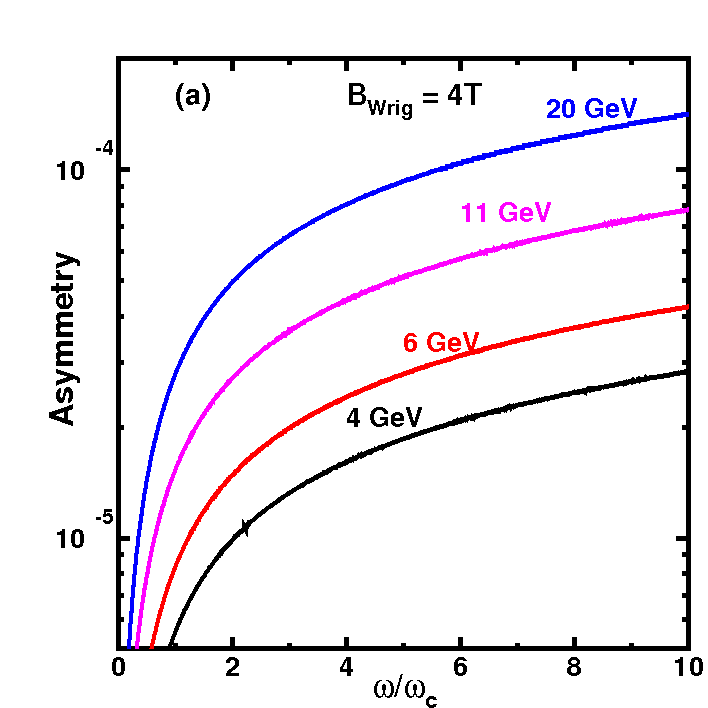}}{\includegraphics*[width=4.5cm]{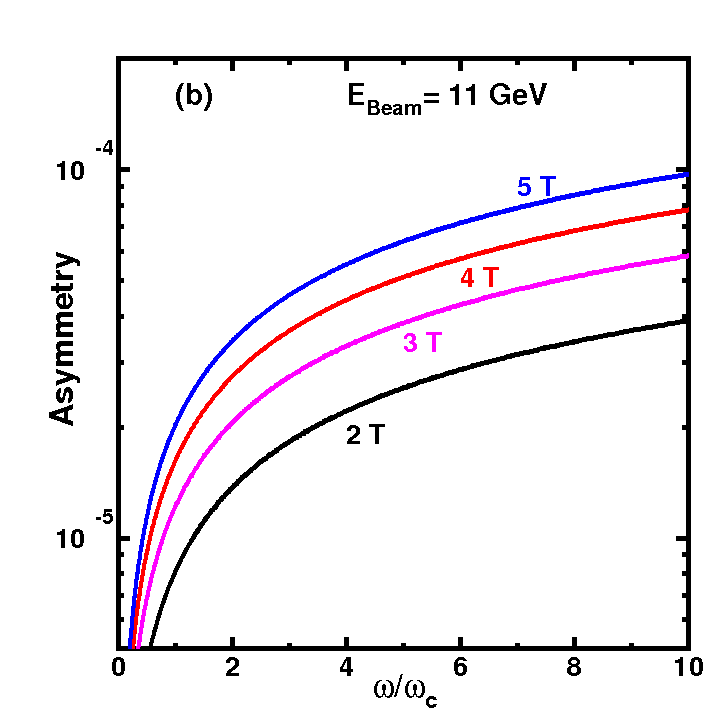}}}
\caption[]{(a) The spin dependent asymmetry as a function of ratio of photon energy to the critical photon energy, for electron beam energy, E$_{beam}$ = 4 -- 20 GeV. (b) The spin dependent asymmetry for magnetic field, B$_{wigg}$ = 2 -- 5 Tesla.}
\label{fig5}
\end{figure}

\begin{figure}[hbt]
\centerline{{\includegraphics*[width=8.0cm]{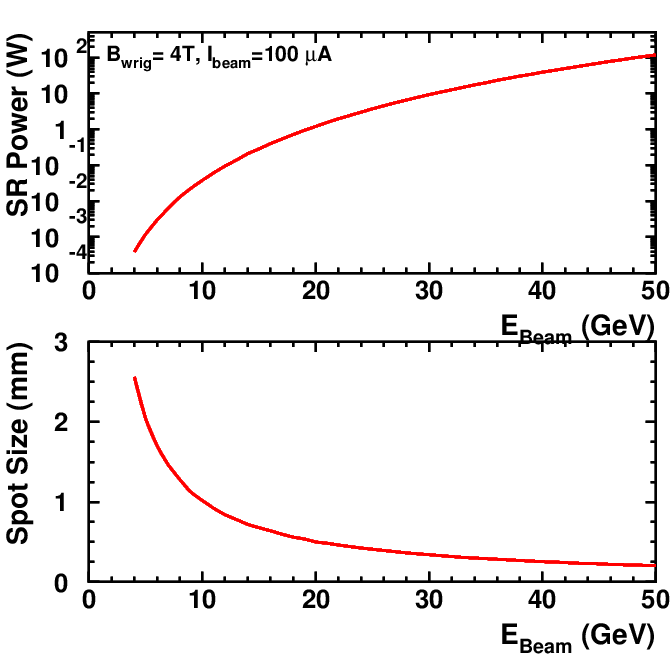}}}
\caption[]{(top) The total SR power radiated as a function of the electron beam energy, for  a 100 $\mu$A beam current and B$_{wigg}$ = 4 Tesla. (bottom) The vertical size of the SR beam spot as a function of the electron beam energy, at a distance of 10~m from the wiggler magnet.}
\label{sr-pow}
\end{figure}
 Fig.~\ref{fig3}~(c) indicates that one should measure the hard tail of the SR spectrum ($E_{\gamma} >$ 500 keV) and avoid the soft part of the spectrum where the asymmetry is low and changes rapidly with energy. Although the asymmetry is small $\sim$ 10$^{-4}$ the photon flux is high, even at the hard tail of the spectrum, allowing a rapid determination of the asymmetry, with 1\% statistical uncertainty within a few tens of seconds ($\frac{\delta A}{A} = \frac{1}{A\sqrt{2N}}$) as shown in Fig~\ref{fig3}~(d). The energy dependence of the asymmetry for $E_e$ = 4 -- 20 GeV and the magnetic field dependence of the asymmetry for $B_{wigg}$ = 2 -- 5 T are shown in Fig~\ref{fig5}(a) and Fig~\ref{fig5}(b) respectively. 
These figures demonstrate that a spin-light based polarimetry is a very promising technique at intermediate energies and can be used to monitor the polarization of 4 -- 20 GeV electrons in very rapid measurement cycles, with high statistical precision.

Although the size of the asymmetry increases with increasing electron beam energy it should be noted that the total power of SR increases as the fourth power of the electron beam energy and increases linearly with beam current. Moreover, the vertical size of the SR spot decreases with increasing electron beam energy as shown in Fig.~\ref{sr-pow} (bottom). These factors impose practical limitations on the 
highest electron beam energies and the highest currents at which a 
spin-light polarimeter would be feasible. We estimate that it is best suited for the 4 - 20 GeV energy range for currents less than $\sim$ 10 mA.

\section{A Conceptual Design}
The two basic components of a spin-light based polarimeter are the source of SR and the X-ray detector which can measure the spatial asymmetry. 

\subsection{The SR Source - Wiggler}
A three pole wiggler magnet with a magnetic field that has uniform magnitude but reversed direction at each pole and a short-long-short pole arrangement is well suited as a source of SR. The three poles must be symmetric about the center such that the line integral of the magnetic field in the direction of the motion of the electron, $z$, must be zero (i.e. $\int B(z)dz$~=~0), ensuring that it does not affect the electron beam transport and its spin direction (beyond the wiggler). The field being  of opposite polarity at the 3 poles, flips the sign of the spin dependent spatial asymmetry from any two adjacent poles and hence when measured simultaneously it can help reduce systematic uncertainties arising from the vertical motion of the beam. 

The intensity and the asymmetry both increase with increasing field strength, while the pole length decreases with increasing field strength. Therefore a field strength of 4~T is a judicious choice for the wiggler field. A 10 mrad bend can be achieved with a pole length of 10~cm. Thus the total magnet length is 40 cm, and the spacing between the poles is optimized for ease of extraction and detection of the SR beam. A separation of 1~m between the poles allows for collimators to be placed that can separate the SR beams spots from the different poles.  The small pole length ensures that the effect of spin-flip inducing SR and the fluctuation of the SR power are negligible ($<$ 0.1\%).

Wiggler magnets are regularly used at light sources around the world such as the Advanced Photon Source (APS) at Argonne National Lab and Spring8 in Japan. Some of these magnets are well suited for a spin-light polarimeter~\cite{wiggler}.

\begin{figure}
\centerline{{\includegraphics*[width=8.0cm]{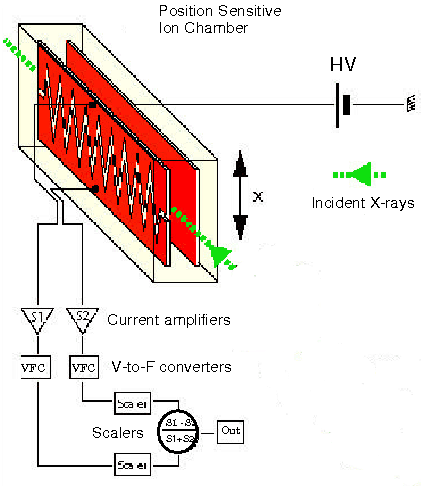}}}
\caption[]{A position sensitive ionization chamber developed at the APS and SPring-8 with a resolution of 5$\mu$m when operated at photon flux of 5.0$\times$10$^{12}$  8 keV photons/sec.~\cite{posic} Note that the image shown here would constitute just one half of a differential ionization chamber required for spin-light polarimetry.}
\label{fig10}
\end{figure}

\subsubsection{Effect of the wiggler on the electron beam}
A non-invasive polarimeter is highly desirable and hence we must study the effect of fluctuations related to the quantum nature of SR produced by the wiggler.
The effect of SR on the electron beam were carefully studied for the recirculating arcs~\cite{norum}, and the same methods can be used to calculate the influence of the wiggler on critical beam parameters. As described in Ref.~\cite{sands} and ~\cite{norum}, the distribution of energies lost by individual electrons in bending through some angle $\theta$ is given by a convolution of the distribution of the number of photons emitted and the distribution of energies of those photons. The number of photons emitted by a particular electron per radian bend will be distributed according to Poisson statistics about a mean value given by ~\cite{sands}, $ n = \frac{5}{2\sqrt{3}}\frac{\gamma}{137} = 20.62 E,$
where $n$ is the mean number of photons per radian bend, and $E$ is the beam energy in GeV. The average energy of the photons emitted is $E_c = \hbar \omega_c = \frac{3}{2}\frac{\hbar c \gamma^3}{R}$~\cite{sands}. Therefore the mean energy fluctuation is given by $\Delta E = \sqrt{n} E_c$. It is interesting to note that the energy fluctuation depends only on the electron beam energy and the bend radius of the wiggler. A beam of 11 GeV electrons in a 4T wiggler with a 10~m bend radius and a bend angle of 10~mrad, gives $n \sim$ 2 and $E_c$ = 199 keV. Therefore $\Delta E/E \sim$ 2.5$\times$10$^{-5}$, which is comparable to the fluctuations due to the recirculating arcs of the JLab accelerator~\cite{norum}.

The SR power spectrum usually peaks at angles of $\pm 1/\gamma$ with respect to the electron direction. However, if an electron emits on the average two photons in a magnet, the angular distribution of the momentum kick received by each electron is peaked in the direction of the electron's motion. The magnitude of the transverse kicks generated by the emission of a photon with energy $E_c$ in the direction $\theta_{\gamma}=1/\gamma$ with respect to the electron direction is given by~\cite{sands}, $\Delta \theta_e = \frac{E_{\gamma}\sin{\theta_{\gamma}}}{E_e} = 11.3\times 10^{-9}\frac{E_e(GeV)}{R(m)}$. The r.m.s. kick from the emission of $n$ photons is given by $\sqrt{n}\Delta \theta_e$. Thus for a 11 GeV beam bend by 10 mrad the r.m.s. kick is $\sim$ 1.5 $\times$ 10$^{-8}$ rad, which is negligible.

Thus the wiggler magnet would have negligible influence on the electron beam and a spin-light polarimeter can be used for non-invasive monitoring of the beam polarization.  

\subsubsection{Influence of the wiggler bend direction}
The wiggler bend direction was chosen to be beam-left for the conceptual design (see for example Figs.~\ref{zoom1} \& ~\ref{zoom2}). Since the bend direction is transverse to the asymmetry direction one does not expect any systematic influence due to the choice of the bend direction. However, the design includes two symmetric pairs of ionization chambers placed on either side of the beam, and since the bend is small (10 mrad), it should be possible to build symmetric pairs of collimators and slits on both sides of the beam. With such a setup the independence of the spin-light asymmetry with respect to the wiggler bend direction can be directly verified during calibration and commissioning of the device. For the stability of operation, changes in bend direction during regular operation is not desirable.   

\subsection{The X-ray Detector - Ionization Chamber}
The detector used to measure the spatial asymmetry must be sensitive to X-rays in the range of about 500 keV to 2.5 MeV and must be able to pick out a small asymmetry from a large spin independent background, it must be radiation hard, have low noise and be able to withstand high rates of $\sim$ 10$^{12}$ photons/sec. Ionization  chambers (IC) are well known for their high rate capability when operated as an integrating detector (i.e. in current mode), low electronic noise and radiation hardness. Argon/Xenon is an attractive candidate for use as an ionization medium, its high atomic number (18/54) and density (when compressed) gives it a high stopping power for hard X-rays and low energy gamma~\cite{specprop}. Over the last two decades, room temperature, high pressure ($>$ 50 atm, 0.55 g/cc) xenon (HPXe) ionization chambers have been developed with high detection efficiency in the 50 keV - 2.0 MeV range~\cite{hpxe1,hpxe2,hpxe3}.
\begin{figure}[hbt]
\centerline{{\includegraphics*[width=8.0cm]{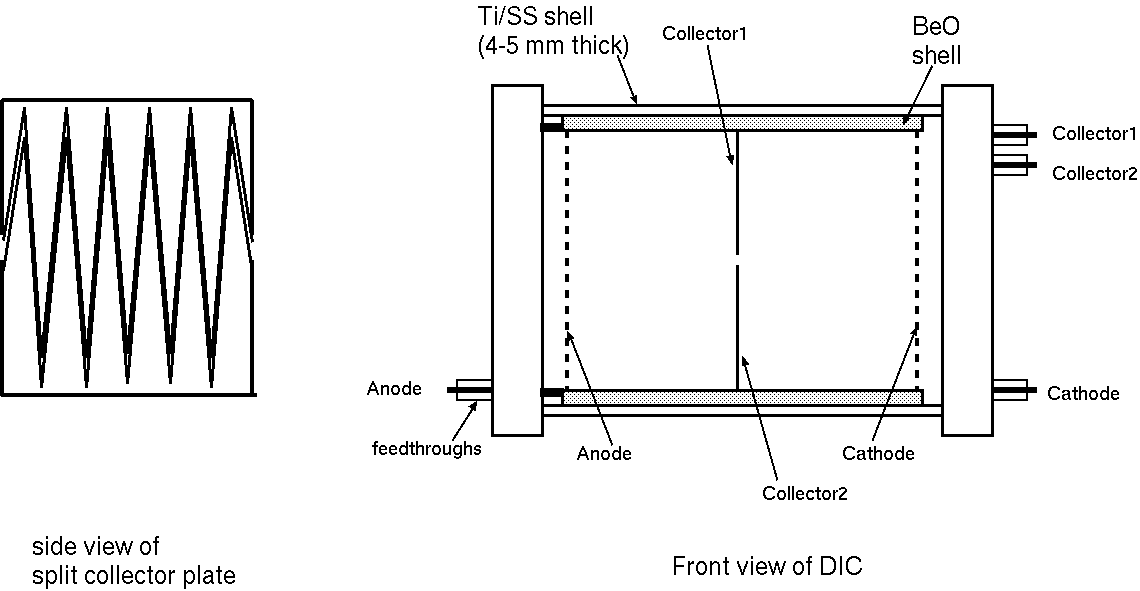}}}
\caption[]{(left) A schematic of the  split anode plate. (right) The dual differential ionization chamber for spin-light polarimetry.}
\label{fig12}
\end{figure}

Another recent development, is the split collector ionization chamber that have turned the IC into a  position sensitive device. Position sensitive ionization chambers are designed to have the collector plate split into two sections in a zig-zag/backgammon pattern such that each half operates as an independent ionization chamber. A prototype of such a chamber has been shown in Fig.~\ref{fig10}. These chambers were developed at the APS at ANL and at the SPring-8 light source in Japan. They are used to measure the vertical position of X-ray beams and have been shown to have a resolution of 5 $\mu$m~\cite{posic}. These chambers also have very low dark currents in the $\sim$pA range and have been operated at photon flux of 5.0$\times$10$^{12}$ photons/sec. They work by measuring the difference in counts between the two halves of the chamber, i.e. they are differential ionization chambers (DIC). A position sensitive DIC operated in current mode can be used to measure the spatial asymmetry of the SR generated by longitudinally polarized electrons. 

A dual, 1 atm. Ar/Xe differential ionization chamber would be ideal for a relative polarimeter. A schematic for such an IC is shown in Fig.~\ref{fig12}. The chamber would consist of Ti or stainless steel windows thick enough to cut down the low energy X-rays ($<$ 50 keV). A pair of split central anode plates (separated by a thin insulator) would be placed between the cathodes. The anode plates would be split in a backgammon pattern. The current measured on each half of the anode plates is amplified with a differential current amplifier.  
\begin{figure}[hbt]
\centerline{{\includegraphics*[width=9.0cm]{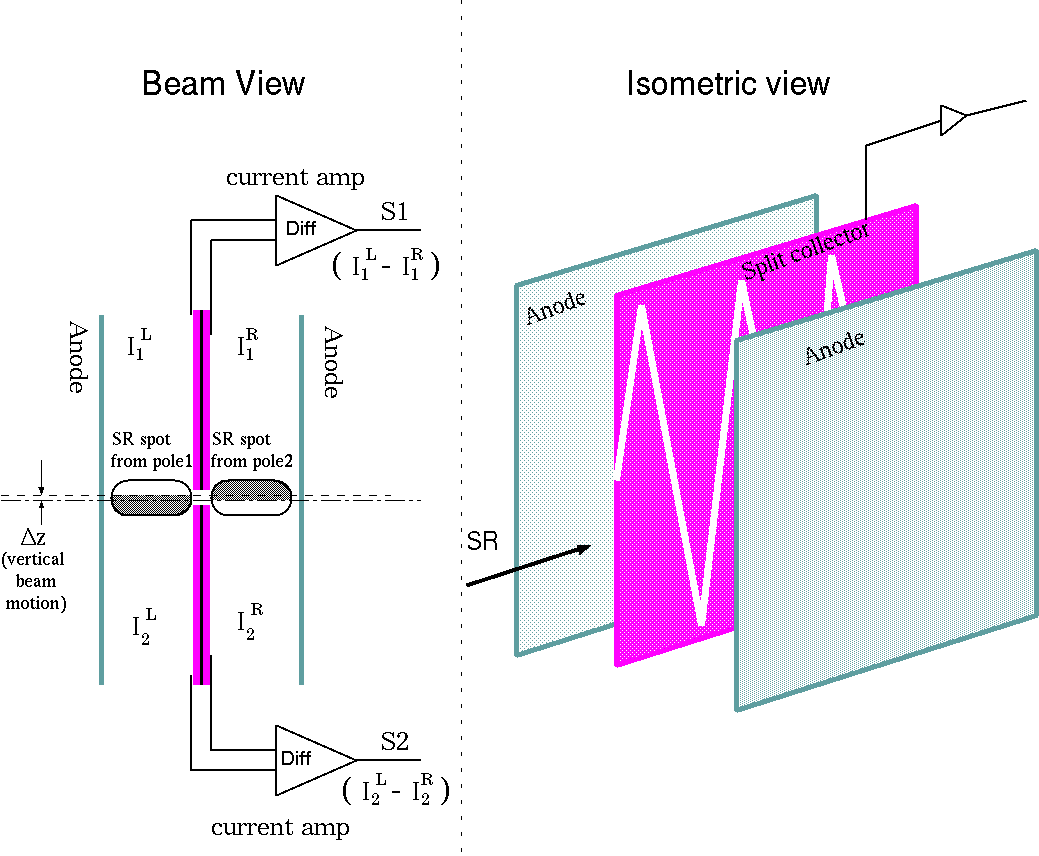}}}
\caption[]{A schematic of the collimated beams of synchrotron radiation from 2 adjacent poles of the wiggler magnet incident on the differential ionization chambers. The figure on the left is a beam's view of the electrodes of the DIC. The collimated radiation is shown as oval blobs, while the figure on the right is an isometric view of the electrodes, without showing the incident radiation.}
\label{diff2}
\end{figure}

A magnified view of the synchrotron radiation from two adjacent poles of the wiggler magnet incident on a dual DIC is shown in Fig.~\ref{diff2}. The left panel of the figure shows the beam's view of the dual DIC. The collimated radiation is shown as oval blobs with the up-down asymmetry represented by the gray shading of the blob (the collimation scheme needed to achieve this is discussed in the next section). The right panel shows an isometric view of the electrodes in the DIC, without showing the incident radiation. The collimated SR beam from two adjacent poles will be incident on opposite sides of the anodes in the dual DIC. The spin-light spatial asymmetry (above and below the orbital plane) will have opposite sign in each half of the DIC  because the magnetic field direction of the adjacent poles of the wiggler are opposite. On the other hand any spatial asymmetry due to vertical motion of the beam will have the same sign in the two halves of the dual DIC and hence should cancel to first order. Thus the dual DIC is essential to ensure that the spin-light polarimeter is insensitive to vertical beam motion.

\subsubsection{The Signal from the DICs}
If we denote $N^{L(R)}_{SR}$ as the number of SR photons on the left(right) of the anode plates, $N^{L(R)}_{spin}$ as the number of spin-light photons, and $\Delta N^{L(R)}_{z}$ as the difference in number of photons introduce by the vertical beam motion, then the contribution to the measured current from the top left part of the dual DIC will be (see Fig.~\ref{diff2}); 
$$I_{1}^{L} \propto N^{L}_{SR}+ N^{L}_{spin}+ \Delta N^{L}_{z},$$ 
similarly the current contribution from the top right of the dual DIC is; 
$$I_{1}^{R} \propto N^{R}_{SR}- N^{R}_{spin}+ \Delta N^{R}_{z}.$$ 
From the simulation studies we estimate that after collimation the size of these currents will be on the order of $\sim$ 10 nA. 

Note the change in the sign of the contribution from spin light photons because they are generated from adjacent poles of the wiggler while the contribution from vertical beam motion has the same sign. For the bottom left and right parts of the dual DIC we get; 
$$I_{2}^{L} \propto N^{L}_{SR}- N^{L}_{spin}- \Delta N^{L}_{z},$$ 
and 
$$I_{2}^{R} \propto N^{R}_{SR}+ N^{R}_{spin}- \Delta N^{R}_{z}.$$ 
Thus, the signal from the top and bottom halves of the DIC, $S1$, and $S2$ as shown in Fig.~\ref{diff2}, can be written as,
\begin{IEEEeqnarray}{rCl}
S1 &\propto& (N^{L}_{SR}+ N^{L}_{spin}+ \Delta N^{L}_{z}) - (N^{R}_{SR} - N^{R}_{spin} + \Delta N^{R}_{z}) \nonumber \\
   & = & 2N_{spin},
\end{IEEEeqnarray} 
and 
\begin{IEEEeqnarray}{rCl}
 S2 &\propto& (N^{L}_{SR}- N^{L}_{spin}- \Delta N^{L}_{z}) - (N^{R}_{SR} + N^{R}_{spin} - \Delta N^{R}_{z}) \nonumber \\
    & = & -2N_{spin}.
\end{IEEEeqnarray}
Hence $S1-S2 \propto 4N_{spin} \propto 4P_{e}$, and the vertical motion related asymmetry cancels to first order as does the corrections due to transverse polarization. However, it should be noted that the signals $S1 + S2$ is proportional to the transverse polarization of the electron beam. This possibility of measuring both the longitudinal and transverse asymmetries in the same setup, provides further capability for reducing systematic uncertainties and makes the spin-light polarimeter an extremely versatile tool.
\begin{figure}[hbt]
\centerline{{\includegraphics*[width=9.0cm]{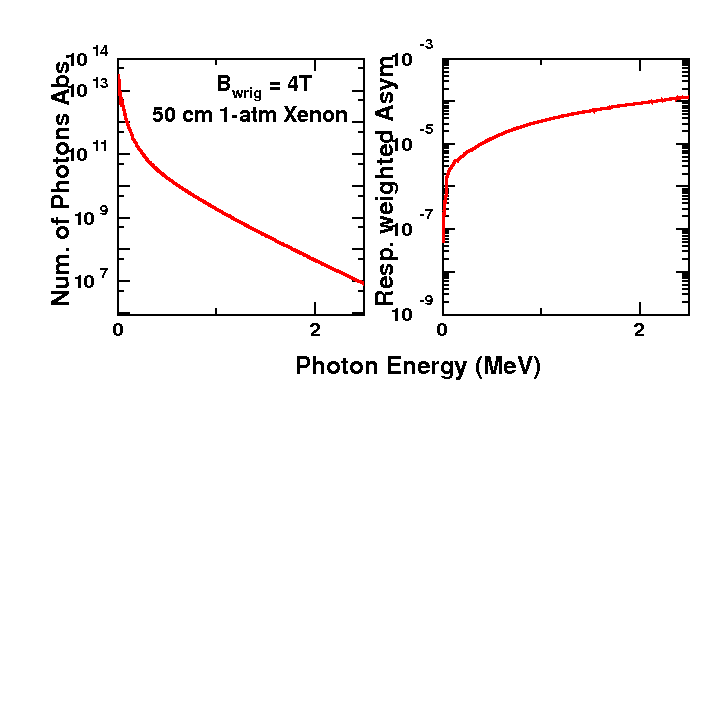}}}
\vspace{-25ex}
\caption[]{(left) The number of photons absorbed in a 1-atm, 50 cm long Xe chamber. (right) The detector response weighted asymmetry.}
\label{abs_coeff}
\end{figure}

The number of photons absorbed in the ionization chamber can be calculated by multiplying the SR spectrum with the absorption function $A(\lambda) = 1 - e^{-\mu(\lambda)\cdot t}$, where $t$ is the length of the chamber, $\lambda$ is the photon wavelength and $\mu$ is the absorption coefficient which is obtained from
NIST database~\cite{coeff}. The number of photons absorbed in a 50 cm long chamber with 1 atm Xe, is shown in Fig.~\ref{abs_coeff}(left). Also shown in the absorption weighted (or detector response weighted) asymmetry (right).

\subsection{Collimation}
\label{sec:coll}
\begin{figure}
\centerline{{\includegraphics*[width=9.0cm]{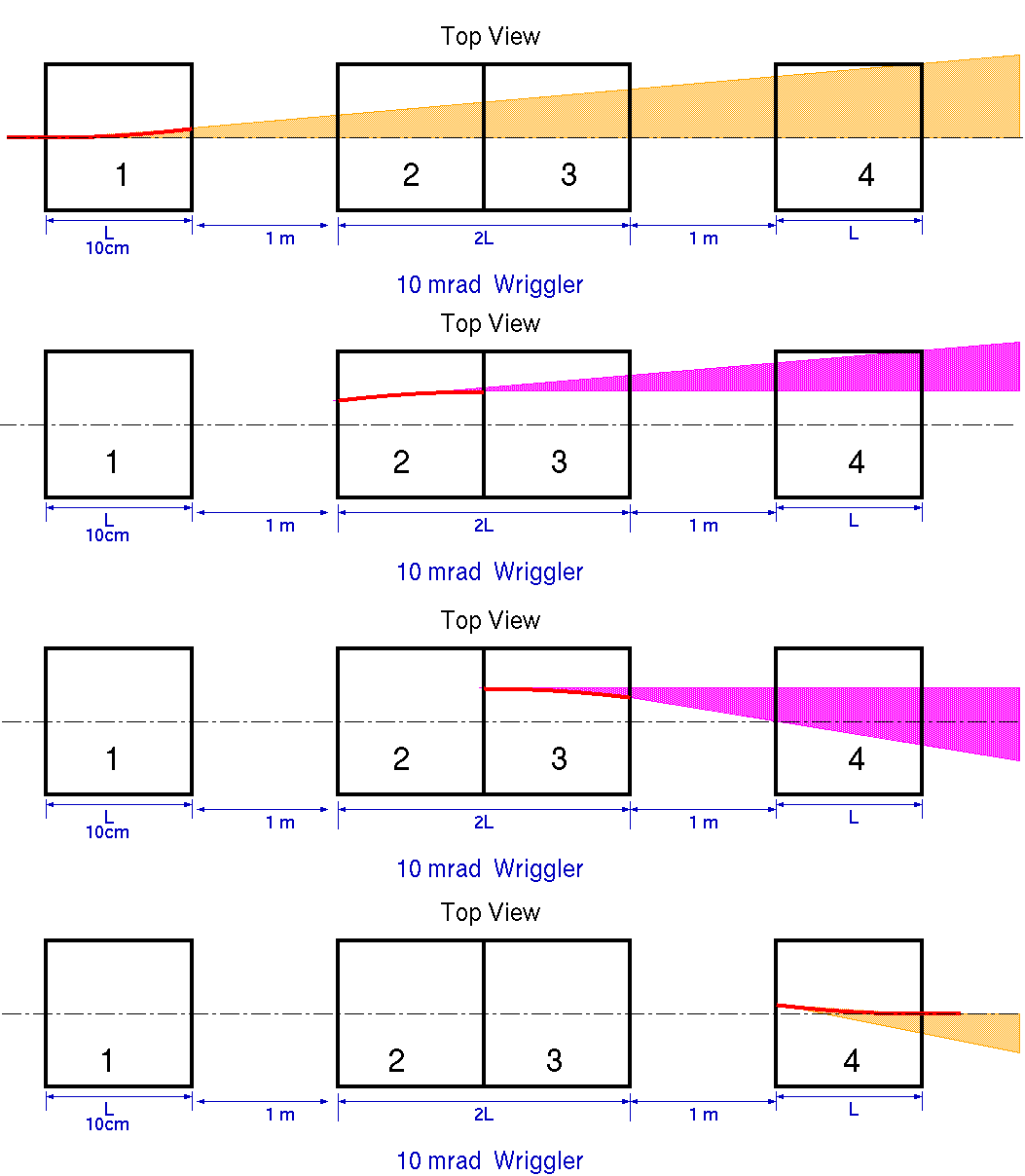}}}
\caption[]{A schematic of the fan of synchrotron radiation produced as the electron beam traverses through each of the 4 poles  of the wiggler magnet. The top view of the magnets has been shown. The two colors are used to indicate that the poles 1 and 4 have opposite polarity compared to poles 2 and 3 and therefore the sign of the asymmetry for the SR fans of the two colors are opposite.}
\label{zoom1}
\end{figure}

\begin{figure}[hbt]
\centerline{{\includegraphics*[width=9.0cm]{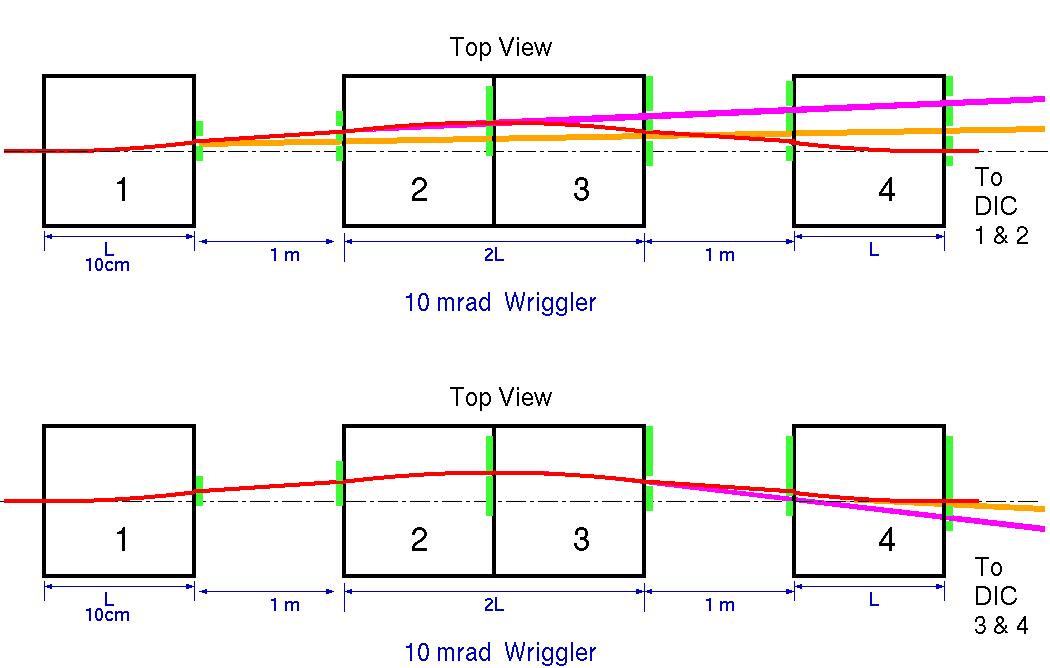}}}
\caption[]{A schematic for the slits and collimators used to select small angular range of the fan of synchrotron radiation. The synchrotron beams from poles 1 and 2 will be incident on the dual DIC placed on the beam left and the beams from poles 3 and 4 are incident on the dual DIC placed on the beam right.}
\label{zoom2}
\end{figure}

The spacing between the wiggler poles was chosen to be 1~m to allow adequate room for the placement of collimators that would separate the SR beams from each
pole. The flight path from the wiggler to the detector is selected to be 10~m, which implies that the SR spot size due to each of the wiggler poles will fan out over a horizontal length of $\sim$ 10~ cm. The vertical width of the SR spot is only $\sim$ 1~mm. With appropriate placement of collimators on the wiggler pole entrance and exit faces, it is possible to separate the SR beam spot from the four different wiggler poles. The magnet system will wiggle the beam by 10 mrad in the horizontal plane such that each pole of the wiggler magnet produces a fan of synchrotron radiation in the horizontal plane as shown in the top view of the magnets (Fig~\ref{zoom1}). 
For longitudinally polarized electrons this fan of synchrotron radiation  will have an up-down asymmetry in the vertical direction (due to spin-light). A series of collimators placed at the front and back faces of each pole of the wiggler magnet and at the center of the central pole of the wiggler will be used to select small angular ranges from the entire fan of synchrotron radiation as shown in the top view of the magnets in Fig~\ref{zoom2}. This collimation scheme enables separation of the synchrotron radiation from each pole of the wiggler magnet. Such a separation is necessary because the up-down asymmetry of the synchrotron radiation has opposite sign for each pole of the wiggler. Each of the collimated beams of synchrotron radiation will be separated by a few cm when they are projected into two symmetric dual DICs located at a distance of 10~m from the wiggler magnet.

\subsection{The Complete Polarimeter}
\begin{figure}[hbt]
\centerline{{\includegraphics*[width=9.0cm]{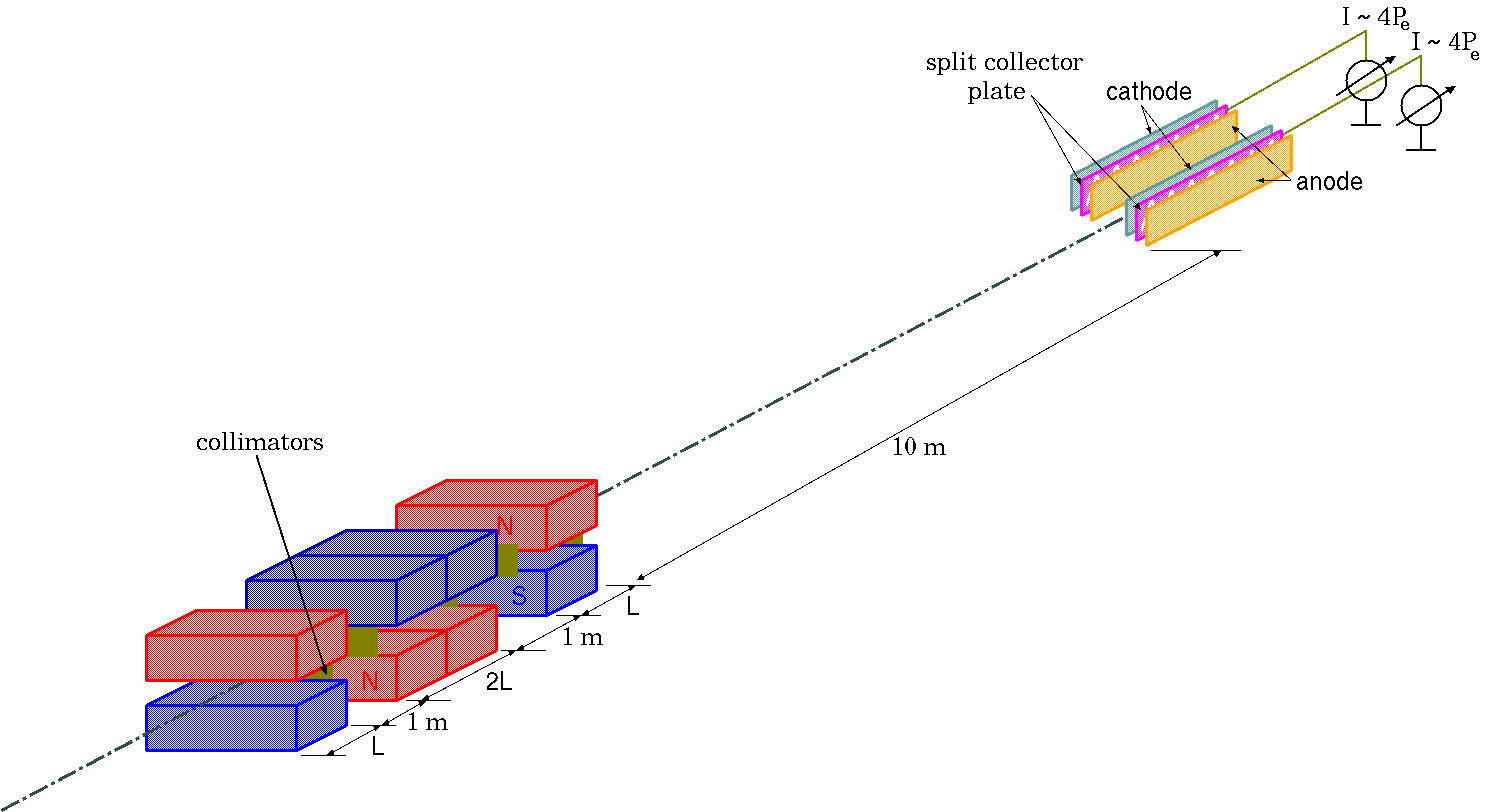}}}
\caption[]{A schematic for a differential spin-light polarimeter (not to scale).}
\label{schematic1}
\end{figure}
A 3D view of the complete spin-light polarimeter is shown in Fig~\ref{schematic1}. Each pole of the wiggler magnet is separated by a distance of 1~m and the two dual differential ionization chambers are placed 10~m from the last pole of wiggler magnet. The vertical backgammon split on the central anodes makes the DIC position sensitive in the vertical direction and hence the signal from the DIC is sensitive to the up-down asymmetry for each of the collimated beams of synchrotron 
radiation. The second dual DIC is necessary to provide an independent measurement of the up-down asymmetry and help reduced systematic uncertainties.  
The main parameters of this conceptual design are tabulated in Table~\ref{tab1}.

\begin{table}
\caption{Parameters of the SR polarimeter at 11 GeV}
\label{tab1}
\begin{center}
\begin{tabular}{|l|l|c|}\hline
&\multicolumn{2}{c}{Beam \& Magnet}\vline \\\hline
E$_e$, current & (GeV), ($\mu$A),& 11.0, 100 \\ \hline
B$_{wiggler}$ &  (T) & 4.0 \\ \hline
Pole (total) length &  (m)& 0.1 (0.4) \\ \hline
Separation & (m) & 1.0 \\
between poles & & \\ \hline
bend angle  &  (mrad) & 10 \\ \hline
vert. opening &  (mrad) & 0.05 \\ 
angle & & \\ \hline
Flight path &(m) &10 \\
to det.     & & \\ \hline
&\multicolumn{2}{c}{SR and detector} \vline \\\hline
 N$_{\gamma}$/s &(Hz)&5.8$\times$ 10$^{13}$ \\ \hline
 $\Delta$N$_{Spin}$/s & (Hz) & 1.8$\times$ 10$^{9}$ \\ \hline
 Detector & (1 atm Xe, cm) & 50 \\ 
medium & &  \\ \hline
N$_{abs}$/s &(Hz) & 3.1$\times$ 10$^{12}$ \\ \hline
$\Delta$N$_{fluctuation}$/s & (Hz) & 7.6$\times$ 10$^6$ \\ \hline
vert. beam & (mm) & 1.0 \\
spot after 10m& & \\
flight& & \\ 
  & &   \\\hline
\end{tabular}
\end{center}
\end{table}

\subsection{Systematic Instrumental Uncertainties}

\begin{figure}[htb]
\centerline{\includegraphics*[width=4.2cm]{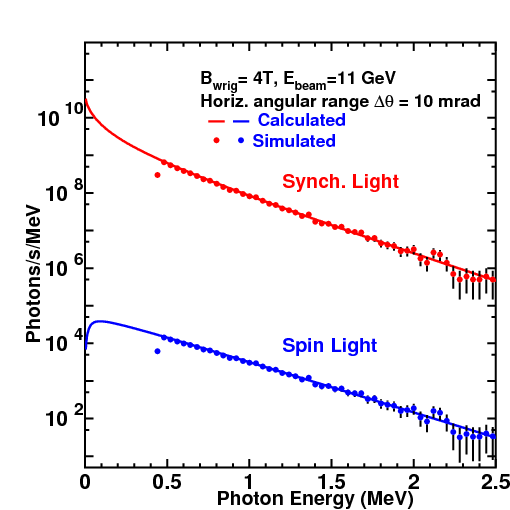} \includegraphics*[width=4.2cm]{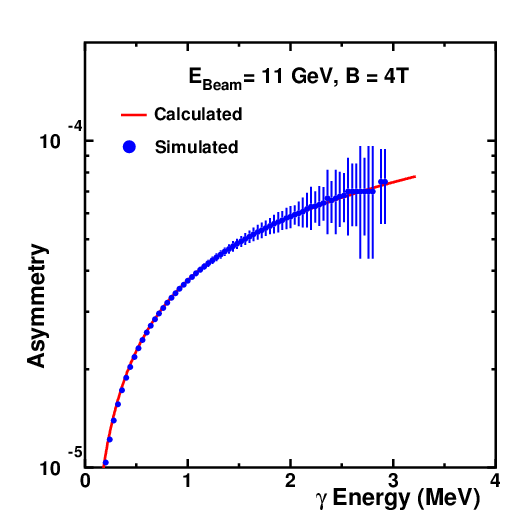}}
\caption[]{The simulated photon spectrum and asymmetry compared to the calculated spectra for the nominal running conditions.}
\label{fig_geant2}
\end{figure}

Some of the major sources of systematic uncertainties for a spin-light polarimeter include the background asymmetries from processes such as Bremsstrahlung and false asymmetry due to vertical beam motion, differences in chamber efficiency and magnetic field non-uniformity between adjacent poles of the wiggler. The measured experimental asymmetry from a spin light polarimeter can be written as $A_{expt} = A_{raw}( 1+ B/S) - A_{B}B/S + A_F$, where $S$ and $B$ are the signal and background, $A_{B}$ is the background asymmetry and $A_{F}$ is the false asymmetry due to factors such as vertical beam motion, differences in the chamber efficiency and differences in the field strength between adjacent poles.  The main advantage of operating the ionization chambers as differential detectors is that the false asymmetries will cancel to first order. In addition the visible portion of the synchrotron light can be used to align the detectors and help control systematic uncertainties. The 3-pole design ensures that the vertical beam motion related false asymmetry also cancels to first order. However, the size of the background must be small compared to the signal. In order to address this issue, a full Geant4~\cite{geant4} simulation of a proto-type spin-light polarimeter was built. The simulation was also used to study the effect of the asymmetry associated with the background. In addition to the built-in synchrotron radiation physics available in Geant4, we have implemented a spin-light generator using a parametrized model of spin-light at the nominal running conditions (11 GeV beam, 4T magnetic field and 10 mrad bend angle). Using this generator we were able to reproduce the expected photon energy spectrum and the expected asymmetry as shown in Fig.~\ref{fig_geant2}.

The exact position and width of the slits/collimators on the front and back faces of the wiggler magnets were optimized iteratively to obtain the best separation between the SR photons originating from the different poles of the magnets. Currently, we are using slits only to illuminate the beam right split ionization chamber. A 3~mm sheet of lead shielding is also applied in front of the ion chamber. 

   
The ionization generated from the photons incident on the ion-chamber is integrated to obtain the signal from the chamber and this signal was used to study several of the systematic uncertainties, such as the effect of background radiation, the position/alignment of the collimators and the alignment of the ion chambers, on the asymmetry signal. The results of these studies have been tabulated in Table~\ref{tab2}.

The simulated data show that the background from non-SR radiation is $\sim$ 1.6\% and most of the background is at energies below 0.5 MeV. If the background related dilution can be determined to $\sim$ 1\% the systematic uncertainty due of background radiation will be  $\sim$ 0.5\%. Moreover, in the proposed setup the background can be determined by measuring the difference in the signal from the chambers with the wiggler magnets turned on and off during calibration and commissioning of the device and at some relatively long interval during regular operations (these background measurements would be invasive in nature). 

The same Geant4 simulation was also used to determine the contribution to the systematic uncertainty due to the uncertainty in the exact position and width of the slits and collimators that are placed on the front and back faces of the poles of the wiggler magnet. These collimators are used to separate the SR photons originating from the various poles. Thus, any uncertainty in the position and width of the collimators can give rise to a false asymmetry due to mixing of spin-light and SR photons from different poles. The simulation demonstrates that a position/width uncertainty of 100 $\mu$m would lead to a uncertainty of $<$ 0.2\%.  

We have also simulated the effects of the finite beam size on the up-down asymmetry. The simulations were performed for a beam with a Gaussian distribution of $\sigma$= 100 $\mu$m. The effects of the fringe fields at the tapered edges of the wiggler poles has also been studied using a Poisson Fish \cite{poissonfish} model of the magnetic field. Although the absolute photon flux was reduced due the fringe fields the effect on the asymmetry was found to be minimal.
 
We have studied the effects of beam halos on the spin-light asymmetry. The restrictive collimation scheme discussed in section~\ref{sec:coll} ensures that the contribution from beam halo is very limited. For a ratio of peak to halo of $\sim~10^8$ the dilution to the spin-light asymmetry is several orders of magnitude smaller than the spin-light asymmetry. Thus, for a tightly controlled beam, halos should not pose a serious challenge.

Most of the systematic uncertainties listed in Table~\ref{tab2} are independent of the electron beam energy in the 4 - 20 GeV range, however, because of the 
decrease in the SR spot size with increasing electron beam energy the uncertainty due the slit width scales inversely with beam energy.   

 The proposed device is best used as a relative polarimeter, however, it can be used as an absolute device if the lower and upper bounds of the energy sensitivity of the DIC is determined accurately. The absolute value of the spin light asymmetry depends on the absolute value of the energy window over which the DIC signals are integrated. It is especially sensitive to the lower bound because of the rapid change in the SR intensity and the spin-light asymmetry as a function of decreasing photon energy. For an absolute measurement the lower bound of the integration window and the sensitivity to the vertical motion of the beam would be the two dominant sources of systematic uncertainty.  Excellent energy resolutions have been demonstrated for HPXe ionization chambers~\cite{hpxe1,hpxe2,hpxe3}. With such high resolution ionization chambers one should be able to determine the response function and the lower bound of energy sensitivity of the chamber to better than 2\%. Using a lower bound of 0.5~MeV, a variation of $\pm$ 10~keV in the lower bound results in a 2.5\% change in the calculated asymmetry. An absolute measurement would be sensitive to vertical beam motion, a $\pm$ 0.1~mm variation in the vertical beam position can result in a 4\% change in the calculated asymmetry.  Thus, a spin-light polarimeter would only 
be capable of $\sim$~5\% absolute polarization measurement.

A table of estimated systematic uncertainties is shown in Table~\ref{tab2}. We estimate the systematic instrumental uncertainties of a relative polarimeter to be $<$ 1\%. 

\begin{table}[hbt]
\caption{Systematic instrumental uncertainties for a relative polarimeter at $E_{beam} = 11$ GeV.}
\label{tab2}
\begin{center}
\begin{tabular}{|l|l|l|}\hline
Source & Uncertainty & $\frac{\delta A}{A}$ \\ \hline
Dark current & $\sim$ pA & $<$ 0.01\% \\ \hline
Intensity fluctuations & $\Delta N \times $ 10$^{-3}$ & $<$0.1 \% \\ \hline
Beam energy & 1.0$\times$10$^{-3}$ & $<$ 0.05 \% \\ \hline
Density of chamber gas, & relative difference  &  $<$0.01\% \\ \hline
Slit width           &  100 $\mu$m  &  $<$0.2 \%        \\ \hline
Background related   & known to 0.5\%        & 0.5 \%        \\ 
dilutions            & for B/S $\sim$ 0.02  &  \\ \hline
Other dilutions      & cancel to first order & $<$ 0.1\% \\\hline
Halo  contributions  & $10^{-8}$  & $<$ 0.1 \% \\ \hline 
Total                &            & 0.6 \% \\ \hline 

\end{tabular}
\end{center}
\end{table}

\section{Conclusion}
Spin light based polarimetry was demonstrated over 30 years ago, but has been ignored since then. A spin-light polarimeter has several advantages over conventional polarimeters and when used in conjunction with a Compton polarimeter it could help provide a new benchmark for precision polarimetry.  The 11 GeV beam at JLab or the electron beam at a future EIC would be well suited for spin light polarimetry and such a polarimeter would help achieve the $<$ 0.5 \% polarimetry desired by experiments approved for the 12 GeV era and proposed for the EIC. A 3 pole wiggler with a field strength of 4~T and a pole length of 10~cm would be adequate for such a polarimeter. A dual position sensitive ionization chambers with split anode plates is ideally suited as the X-ray detector for such a polarimeter. The differential detector design would help reduce systematic uncertainties.


%

\section*{Acknowledgment}
This work was supported in part by the U.S. Department of Energy under contract \# DE-FG02-07ER41528, by the EIC Detector R\&D grant from Brookhaven National Lab.
One of us (P.M.) would also like the thank the Jefferson Science Associates for a JSA Fellowship.

\ifCLASSOPTIONcaptionsoff
  \newpage
\fi




\begin{thebibliography}{1}
\bibitem{hallcmoller}M. Hauger {\it et al.}, Nucl. Inst. Meth. {\bf A462}, 382 (2001).
\bibitem{qweak} To be reported for the polarization measured during the QWeak experiment at JLab, Hall-C.
\bibitem{sld}P. C. Rowson, D. Su, and S. Willocq, Ann. Rev. Nucl. Part. Sci. {\bf 51}, 345 (2001); M. Woods, SLAC-PUB-7319 (1996).
\bibitem{geant4}J. Allison {\it et al.}, Nucl. Inst. and Meth. {\bf A506}, 250 (2003); J. Allison {\it et al.}, IEEE Trans. in Nucl. Science {\bf 53}, 270 (2006).
\bibitem{karabekov93}I.~P.~Karabekov, R.~Rossmanith, Proc. of the 1993 PAC, Washington, v.~1, p. 457 (1993); I.~P.~Karabekov and S.~I.~Karabekian, Proceedings of 5th European Particle Accelerator Conference (EPAC 96), Sitges, Spain, 10-14 Jun 1996, pp 1743-1745 (1996); A.~V.~Airapetian, R.~O.~Avakian, I.~P.~Karabekov, E.~L.~Saldin, and M.~V.~Yurkov, Proc. of the SPIN-96, Amsterdam, The Netherlands, Vol.1, p762 (1996). 
\bibitem{classical}D.~D.~Ivanenko, I. Pomeranchuk, Ya Zh. Eksp. Teor. Fiz. {\bf 16}, 370 (1946); J.~Schwinger, Phys. Rev. {\bf 75}, 1912 (1947).
\bibitem{polar}G.~A.~Schott, Ann. Phys. {\bf 24}, 635 (1907); A.~A.~Sokolov and I.~M.~Ternov, Zh. Eksp. Theor. Fiz. {\bf 31}, 373 (1956), Sov. Phys. JETP {\bf 4}, 396 (1957). 
\bibitem{sync1}F.~R. Elder, R. V. Langmuir and H. C. Pollock, Phys. Rev. {\bf 74}, 52 (1948).
\bibitem{sync2}M. Yu Ado and P.~A.~ Cherenkov, Sov. Phys. Dokl. {\bf 1}, 517 (1957).
\bibitem{sync3}F.~A.~Korolev, E.~.N.~Akimov, E.~N.~Markov, and O.~F.~Kulikov Sov. Phys. Dokl. {\bf 1}, 568 (1957).
\bibitem{sync4}P.~Joos, Phys. Rev. Lett. {\bf 4}, 558 (1960).
\bibitem{stk52}A.~A.~Sokolov, N.~P.~Klepikov and I.~M.~Ternov, JETF {\bf 23}, 632 (1952).
\bibitem{sokolov53}A.~A.~Sokolov, and I.~.M.~Ternov, JETF {\bf 25}, 698 (1953).
\bibitem{sttext}A.~A.~ Sokolov and I.~.M.~Ternov, {\it Synchrotron Radiation}, Pergamon Press, New York (1968); A.~A.~ Sokolov and I.~.M.~Ternov, {\it Radiation from Relativistic Electrons}, A.I.P. Translation Series, New York (1986).
\bibitem{ternov95} I. M. Ternov, Physics - Uspekhi {\bf 38}, 409 (1995).
\bibitem{bordo}V. A. Bordovitsyn, Ph. D. Thesis, Moscow (1983); I. M. Ternov and V. A. Bordovitsyn, Vestn. Mosk. Univ. Ser. Fiz. Astr. {\bf 24}, 69 (1983); V. A. Bordovitsyn and V. V. Telushkin, Nucl. Inst. and Meth. {\bf B266}, 3708 (2008).
\bibitem{belom82} S.~A.~Belomesthnykh {\it et al.}, Nucl. Inst. and Meth. {\bf 227}, 173 (1984).
\bibitem{marin} J.~ Le Duff, P.~C.~Marin, J.~L.~Manson, and M.~Sommev, Orsay - Rapport Technique, 4-73 (1973).
\bibitem{hpxe1}V.~V.~Dmitrenko {\it et al.}, Sov. Phys.-tech. Phys. {\bf 28}, 1440 (1983); A.~E.~Bolotnikov {\it et al.}, Sov. Phys.-Tech. Phys. {\bf 33}, 449 (1988); 
\bibitem{hpxe2}C.~Levin {\it et al.}, Nucl. Inst. and Meth. {\bf A332}, 206 (1993). 
\bibitem{hpxe3}G. Tepper and J. Losee, Nucl. Inst. and Meth. {\bf A356}, 339 (1995). 
\bibitem{wiggler}E. Nakamura {\it et al.}, J. of Elec. Spec. and Rel. Phen. {\bf 80}, 421 (1996); D.~E.~Baynham, P. T. M. Clee, and D. J. Thompson, Nucl. Instr. and Meth., {\bf 152}, 31 (1978)
\bibitem{posic} K. Sato, J. of Synchrotron Rad., {\bf 8}, 378 (2001); T. Gog, D.~M. ~Casa and I. Kuzmenko, CMC-CAT technical report.
\bibitem{sands} M. Sands, SLAC Technical note, SLAC-121 (1970). 
\bibitem{norum} B. Norum, CEBAF Technical note, TN-0019 (1985).
\bibitem{specprop} A.~E.~Bolotnikov and B.~Ramsey, Nucl. Inst. and Meth. {\bf A396}, 360 (1997).
\bibitem{coeff}http://www.nist.gov/physlab/data/xraycoef/index.cfm
\bibitem{poissonfish}http://library.lanl.gov/cgi-bin/getfile?00415886.pdf
\end{thebibliography}
%

%


\begin{IEEEbiographynophoto}{Prajwal Mohanmurthy}
Prajwal Mohanmurthy obtained his Bachelor of Science degree from Mississippi State University in 2012. He was a graduate research fellow at the High Performance Computing Collaboratory at Mississippi State University and currently is a graduate research fellow in the Laboratory for Nuclear Sciences at Massachusetts Institute of Technology. His research interests are centered around test of standard model and fundamental symmetries in search of physics beyond the standard model. His recent research involvements have been geared towards a search for Axionic Dark Matter and the precision measurement of the mass of neutrinos. He also actively collaborates to develop beam instrumentation for up and coming facilities and future accelerators.
\end{IEEEbiographynophoto}


\begin{IEEEbiographynophoto}{Dipangkar Dutta}
Dr. Dipangkar Dutta is an Associate Professor of Physics at the Mississippi State University Department of Physics and Astronomy. He obtained his Bachelor of Technology degree from Indian Institute of Technology, Bombay in 1992 and his doctoral degree in Physics from Northwestern University in 1999. He was a post-doctoral and senior post-doctoral fellow in the Laboratory for Nuclear Sciences at Massachusetts Institute of Technology. His research is focused primarily on precision measurement of fundamental properties of nucleons. He is also interested in precision tests of fundamental symmetries and the Standard Model.
\end{IEEEbiographynophoto}




\end{document}